\documentclass{desyprocA4j}

\begin{document}
\title{Long-lived charged sleptons at the ILC/CLIC}

\author{{\slshape Jan Heisig}\\[1ex]
II.~Institute for Theoretical Physics, University of Hamburg,
Germany\\[3ex]
LC-REP-2012-065}

\maketitle 

\begin{abstract}
Supersymmetric scenarios with a very weakly interacting lightest 
superpartner (LSP)---like the gravitino or axino---naturally give 
rise to a long-lived next-to-LSP (NLSP). 
If the NLSP is a charged slepton it leaves a very distinct signature 
in a collider experiment. At the ILC/CLIC it will be possible to capture 
a significant fraction of the produced charged sleptons and
observe their decays. These decays potentially reveal the nature of 
the LSP and thus provide a unique possibility to measure the properties 
of a very weakly interacting LSP which otherwise is most likely 
hidden from any other observation, like direct or indirect dark 
matter searches. We review the proposals that have been made to
measure the LSP properties at the ILC/CLIC and compare its potential
to the capability of the LHC\@.
\end{abstract}

\section{Introduction}

In supersymmetric extensions of the standard model (SM) with conserved 
$R$-parity the lightest superpartner (LSP) is stable and thus provides a 
natural dark matter (DM) candidate. The lightest neutralino---being part of 
the minimal supersymmetric standard model (MSSM)---is the most widely 
studied candidate. 
However, in extensions of the MSSM other cosmologically viable DM 
candidates appear such as the gravitino or the axino.

The spin-$3/2$ gravitino $\widetilde G$ arises in the spectrum of supergravity, 
i.e., once supersymmetry (SUSY) is promoted from a global to a local symmetry. 
It is a well motivated DM candidate and can even be regarded
as favored since it alleviates the cosmological gravitino problem \cite{Ellis:1984er}
allowing for a higher reheating temperature as required for thermal leptogenesis 
\cite{Bolz:1998ek}. The gravitino acquires a mass through the super Higgs 
mechanism once SUSY is broken. Its mass depends strongly on the SUSY breaking
scheme and can range from the eV scale to scales beyond the TeV scale. 
Requiring a reheating temperature of $\mathcal{O} (10^9\,$GeV$)$,
masses of around and above $\mathcal{O} (10\,$GeV$)$ are favored in order not to
over-close the universe by the thermally produced gravitino abundance.
The very weak coupling of the gravitino causes the next-to-LSP (NLSP) to be
long-lived. Thus, in the early universe after the NLSP freeze-out, 
late NLSP decays taking place during or after big bang nucleosynthesis (BBN) 
can affect the primordial abundance of light elements. This imposes strong constraints 
on the couplings and lifetime of the NLSP.
Accordingly, a neutralino NLSP is strongly disfavored by BBN constraints from energy 
injection \cite{Feng:2004mt,Roszkowski:2004jd,Cerdeno:2005eu,Cyburt:2006uv}. 
The lighter stau $\widetilde\tau_1$ is therefore often considered 
as a natural NLSP candidate.\footnote{%
The basic ideas given in this article are expected to hold with modifications 
for other charged NLSP candidates or even for very different scenarios. 
Some of the ideas discussed in this article have initially been 
brought up in the context of 4th generation lepton searches \cite{Goity:1993ih}.}
The most severe bound on the stau NLSP lifetime arises from $^6$Li/H constraints
requiring $\tau_{\widetilde \tau_1}\lesssim 5\times10^3\,$s 
\cite{Pospelov:2006sc,Pradler:2007is} for a typical stau yield after freeze out. 
The most conservative bound arises from $^3$He/D constraint. 
It excludes lifetimes $\tau_{\widetilde \tau_1}\gtrsim 10^6\,$s \cite{Pradler:2007ar}.
Conclusively, lifetimes ranging from seconds to a month may be considered 
as interesting.

The resulting signatures of staus at colliders are charged, muon-like tracks 
usually leaving the detector---the decay length is large compared to the
size of a detector. The tracks of staus can be discriminated against the muon 
background via high ionization loss and anomalous time-of-flight. 
The LHC provides a good environment for discovering long-lived staus. 
Searches for heavy stable charged particles are being performed at ATLAS 
\cite{Aad:2011hz} and CMS \cite{Chatrchyan1205.0272}. 

Ionization loss is the main source of energy loss for heavy charged particles
when penetrating the detector material. The energy loss increases with
decreasing velocity $\beta$. Staus that are produced with sufficiently 
small $\beta$ may lose their kinetic energy completely and stop inside 
the detector. According to its lifetime, the stau will decay leaving a characteristic 
signature in the detector which is uncorrelated with the bunch crossing. 
If it is possible to measure the lifetime, the recoil energy and even the angular 
distribution of the emitted SM particles in the decay, it is possible to determine 
the coupling, mass and even the spin of the LSP.
This is a unique possibility to test a (stable) gravitino DM scenario which
is hopeless to test in direct and indirect DM searches.

Another well motivated DM candidate is the axino $\widetilde a$ which appears 
once the MSSM is extended by the Peccei-Quinn mechanism, in order to solve 
the strong CP problem.
The phenomenology at a collider is virtually identical. 
The decay of the stau into the axino can give insights into the Peccei-Quinn sector.

We will consider both scenarios here.
In section \ref{sec:decays} we will describe the decays of the NLSP into
the gravitino or axino LSP and explain how to distinguish these cases. 
In section \ref{sec:LHC} we will describe the implications from the LHC and
its sensitivity to these scenarios. In section \ref{sec:LC} we will review some of the
experimental ideas that have been brought up in order to realize the investigation 
of NLSP decays.

\section{NLSP decays} \label{sec:decays}

In the considered scenarios the dominant decay mode of the staus 
is the 2-body decay $\widetilde \tau \to\widetilde G \tau$ or 
$\widetilde \tau \to\widetilde a \tau$. 
For the gravitino LSP the corresponding decay width reads 
\begin{equation}
\Gamma(\widetilde \tau_1 \to\widetilde G \tau)\simeq
\frac{m_{\widetilde\tau_1}^5}{48\pi \,m_{\widetilde G}^2 M_{\text{Pl}}^2}
\left(1-\frac{m_{\widetilde G}^2}{m_{\widetilde\tau_1}^2}\right)^4\,,
\label{eq:decay-gravitino}
\end{equation}
where $M_{\text{Pl}}$ is the (reduced) Planck mass.
The decay rate is completely determined by the masses $m_{\widetilde\tau_1}$
and $m_{\widetilde G}$. It is independent of any other SUSY parameter or SM 
coupling.

For the axino LSP the 2-body decay is loop-induced and contains further 
SUSY parameters in particular it depends on the stau mixing angle. For a
pure right-handed stau the width has been computed in the KSVZ axino model
\cite{Brandenburg:2005he},
\begin{equation}
\Gamma(\widetilde \tau_1 \to\widetilde a \tau)\simeq
\frac{9\alpha^4C_{\text{aYY}}^2}{512\pi^5\cos^8\theta_{\text{W}}}
\frac{m_{\widetilde B}^2}{f_a^2}
\frac{(m_{\widetilde\tau_1}^2-m_{\widetilde a}^2)^2}{m_{\widetilde\tau_1}^3}
\xi^2\log^2\left(\frac{f_a}{m_{\widetilde\tau_1}}\right)\,,
\label{eq:decay-axino}
\end{equation}
where $\alpha$ is the fine structure constant, $\theta_{\text{W}}$ is the weak mixing
angle, $f_a$ is the Peccei-Quinn scale, $m_{\widetilde B}$ is the (pure) bino
mass and $C_{\text{aYY}}$ and $\xi$ are $\mathcal{O}(1)$ factors expressing
the Peccei-Quinn model dependence and loop cut-off uncertainties, respectively.

The typical decay length of the staus is large compared to their traveling range 
in the detector material. Hence, staus always decay at rest, i.e.,
we know the center-of-mass frame. Accordingly, if the mass of the 
stau is known, the LSP mass can be determined from
the recoil energy of the $\tau$ produced in the 2-body decay, $E_\tau$,
\begin{equation}
m_{\text{LSP}}=\sqrt{m_{\widetilde\tau_1}^2 + m_\tau^2 - 2m_{\widetilde\tau_1}E_\tau}\,.
\label{eq:massreco}
\end{equation}
As pointed out in \cite{Buchmuller:2004rq,Feng:2004gn},
we can probe the hypothesis of a gravitino LSP by computing the
Planck mass from (\ref{eq:decay-gravitino}) once $m_{\widetilde\tau_1}$, 
$m_{\text{LSP}}$ and lifetime $\tau_{\widetilde\tau_1}=\Gamma_{\widetilde\tau_1}^{-1}$ 
are known. An agreement with the Planck mass measured in macroscopic experiments 
would provide a strong evidence for supergravity and the existence of the gravitino.
Since the gravitino mass is directly related to the scale of spontaneous 
SUSY breaking,
\begin{equation}
\langle F \rangle=\sqrt{3}M_{\text{Pl}}\,m_{\widetilde G}\,,
\end{equation}
these measurements would provide us with insights in the SUSY 
breaking sector that are otherwise beyond the reach of any experiment
in the near future.
For the axino LSP case, from (\ref{eq:decay-axino}) we may be able to
estimate the Peccei-Quinn scale and confront it with limits from
astrophysical axion studies and axion searches in the laboratory. 

A sub-dominant but nevertheless very important decay mode of the stau
is the 3-body decay $\widetilde \tau \to\widetilde G \tau\gamma$ or
$\widetilde \tau \to\widetilde a \tau\gamma$ which 
has been studied in \cite{Buchmuller:2004rq,Brandenburg:2005he}.
As pointed out in these references, from the 3-body decay branching ratio
as well as from the distribution of the angle between the $\tau$ and the photon, 
the spin of the LSP can be determined. 
More precisely, it has been shown that it is possible to
distinguish between the spin-$3/2$ gravitino and a spin-$1/2$ axino. 
The observation of a spin-$3/2$ LSP would be an important 
confirmation of supergravity. In particular, for small gravitino masses 
$m_{\widetilde G}\lesssim 0.1\, m_{\widetilde\tau_1}$ the determination
of $m_{\widetilde G}$ requires a very precise measurement of the
tau recoil energy at below the percent level. Thus, (\ref{eq:massreco})
may only provide an upper limit on the gravitino mass in these cases. 
In such a situation a much better determination of $m_{\widetilde G}$ 
can be achieved via (\ref{eq:decay-gravitino}) from the measurement 
of the stau lifetime once we are convinced that the LSP 
is indeed a gravitino by the measurement of its spin.

\section{Implications from the LHC} \label{sec:LHC}

Before the stau will be observed at the ILC/CLIC we
expect its discovery at the LHC.\footnote{%
In the long-lived stau scenario there are very little regions in 
parameter space that are not accessible with the long-term
14\,TeV LHC run but with a mid-term 3\,TeV CLIC run.}
Therefore, in this section we will briefly review the LHC
potential.

Long-lived staus leave a prominent signature in the detectors
of the LHC. Combining ionization loss and time-of-flight measurements
provide very clean signal regions and, at the same time, high efficiencies.
Consequently, the discovery of long-lived staus typically can be claimed on the 
basis of a very few events and thus is expected to be established in a rather 
short time period without providing any hints in advance.

\begin{figure}[tbhp]
\begin{center}
\begin{picture}(400,175)
\put(0,-10){\includegraphics[scale=1]{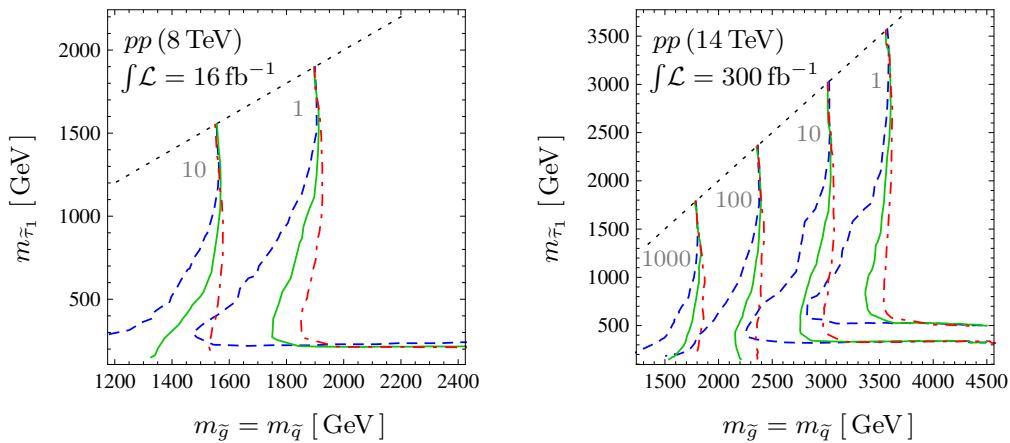}}
\end{picture}
\end{center}
\caption{
Expected number of events that contain staus that are stopped inside
an LHC detector. The results are expressed in a simplified model
framework considering direct stau production
as well as the production via the decay of strongly produced sparticles. 
A common squark and gluino mass, $m_{\widetilde q}=m_{\widetilde g}$,
has been chosen. 
The three different line styles refer to three different mass patterns
of intermediate sparticles in the decay chain. Taken from \cite{Heisig:2012zq}. 
}
\label{fig:LHC}
\end{figure}

The direct production of staus provides a robust lower limit on the stau
mass \cite{Heisig:2011dr,Lindert:2011td}. Null searches for this channel at the 
7\,TeV, 5\,fb$^{-1}$ LHC run \cite{Chatrchyan1205.0272}
can be interpreted in the most conservative limit to exclude 
stau masses below $216\,$GeV \cite{Heisig:2012ep}.
Although the LHC provides a very good environment to discover
heavy stable charged particles, it is typically difficult to capture a 
sufficiently large number of staus in the detector in 
order to be able to study its decays systematically. 
As shown in \cite{Heisig:2012zq}
especially widely spread spectra (spectra with large mass gaps between 
the colored sparticles and the stau) provide way too little stopped staus
for the desired measurements (see figure \ref{fig:LHC}). 
For such spectra even a scenario with 
$m_{\widetilde\tau_1}$ just above the above quoted limit provides less than
100 events of staus that are stopped inside a LHC
detector for the 14\,TeV, 300\,fb$^{-1}$ LHC run.
Proposals to study stopped staus at the LHC are discussed in 
\cite{Hamaguchi:2006vu,Pinfold:2010aq,Graham:2011ah,Feng:2004yi,Hamaguchi:2004df,DeRoeck:2005bw}.

\section{Prospects at the ILC/CLIC} \label{sec:LC}

The challenge in the study of stau decays is to trap as many staus as possible in a 
well defined volume that is sensitive to the observables of the produced SM
particles in the decay. An $e^+e^-$-collider provides an appropriate environment for
this task. On the one hand the direct production of staus provides a
velocity distribution that can be tuned through the center-of-mass energy 
in order to maximize the number of stopped staus in a given volume. 
On the other hand it provides a well defined angular distribution. 
Together with the option of adding extra stopping material  in 
appropriate regions \cite{Hamaguchi:2004df} it provides an ideal framework 
to obtain a large number of observed stau decays. 

The stau may be produced directly or in a decay chain following the production 
of other sparticles. The cross sections for different production processes have different
velocity dependencies near threshold. For slepton production via $s$-channel 
$\gamma/Z$ the cross section increases as $\beta^3$. 
For polarized $e^+e^-$ beams the production cross section for 
selectron pairs via $t$-channel $\widetilde\chi^0$ exchange
($e_{\text{L}}^+e_{\text{L}}^- \to \widetilde e_{\text{R}}^+ \widetilde e_{\text{L}}^-$ 
or $e_{\text{R}}^+e_{\text{R}}^- \to \widetilde e_{\text{L}}^+ \widetilde e_{\text{R}}^-$)
increases linear in $\beta$ and thus provides an enhanced number of selectrons 
close to threshold \cite{Freitas:2001zh,Freitas:2003yp}.\footnote{%
In \cite{Feng:1998ud,Feng:2001ce} the possibility of an $e^-e^-$-collider
to obtain a $\,\propto\!\beta$-behavior near threshold has been discussed.}
Hence, if the spectrum features a selectron which is close in mass to the stau,
one could greatly benefit from the use of polarized electron beams to increase 
the number of produced selectrons near threshold and therefore increase the 
number of stopped staus. For small mass gaps between the selectron and the 
stau this advantage overcompensates the boost that staus achieve from the 
decay of the selectron (which would lead to higher stau velocities).

Once a stau pair is produced it will be identified via highly ionizing tracks. 
Their passage through the detector can be accurately followed.
If the stau stops inside the detector the location of the stopped stau is expected to be
determinable within a volume of a few cm$^3$ \cite{Martyn:2006as}.
The location and time of the stopped stau may be recorded. 
In general the stau  will decay out-of-time with the beam collisions.
Hence, the decay can then be triggered by an isolated, out-of-time hadronic 
or electromagnetic cluster in the hadronic calorimeter (HCAL), 
a hadronic shower in the iron yoke or by a muon originating in the HCAL 
or yoke above an appropriate energy threshold ($E>10\,$GeV) \cite{Martyn:2006as}. 
Background from cosmic rays may be rejected by a veto against vertices in the 
outermost detector layers.
Background from atmospheric neutrinos is expected to be sufficiently rejected
by the required energy threshold and furthermore by the requirement of a matching 
of the recorded stopping positions \cite{Hamaguchi:2004df}.
A precise measurement of the stau mass which is required in order to estimate 
the gravitino mass can be obtained from the reconstruction of the complete event
kinematics.

The potential to measure $m_{\widetilde\tau_1}$, $m_{\widetilde G}$ and
$\tau_{\widetilde\tau_1}$ at the ILC/CLIC equipped with a general purpose detector
\cite{Linssen:2012hp,BrauJames:2007aa}
has been studied for several benchmark points in \cite{Martyn:2006as,Cakir:2007xa}.
Both studies contain the mSUGRA points GDM $\zeta$ 
($m_{\widetilde\tau_1}=346$\,GeV, $m_{\widetilde G}=100\,$GeV)
and GDM $\eta$ ($m_{\widetilde\tau_1}=327\,$GeV, $m_{\widetilde G}=20\,$GeV) 
\cite{DeRoeck:2005bw}. Provided a fixed center-of-mass energy of 800\,GeV 
and a luminosity of 1000\,fb$^{-1}$, $m_{\widetilde\tau_1}$ and 
$\tau_{\widetilde\tau_1}$ have been found to be measurable at the level of one 
per mille and a few per cent, respectively, for both scenarios. 
The gravitino mass $m_{\widetilde G}$ has been found to be measurable at a ten 
per cent level for GDM $\zeta$ and with an uncertainty comparable to its actual value 
for GDM $\eta$ \cite{Martyn:2006as}. These numbers have been obtained with
unpolarized beams. Polarization is expected to enhance the number of stopped 
staus by a factor of almost three \cite{Martyn:2006as} and thus improve these results.
The optimization of the beam energy for given stau masses and production processes 
has been discussed in \cite{Cakir:2007xa}.

Further optimizations can be achieved by placing additional active stopper 
material \cite{Hamaguchi:2004df} around the general purpose detector. Another
approach is the installation of water tanks \cite{Feng:2004yi} that accumulate stopped 
staus. The water can then be transported to a quiet environment in order to study the
decays. It has also been proposed to collect staus in a storage ring 
\cite{Buchmuller:2004rq}.
This could most easily been done if staus where produced preferably in the forward
region, i.e., via selectron pair production (see figure 11 in \cite{Cakir:2007xa}).

\begin{figure}[tbhp]
\begin{center}
\begin{picture}(400,195)
\put(-10,-10){\includegraphics[scale=0.66]{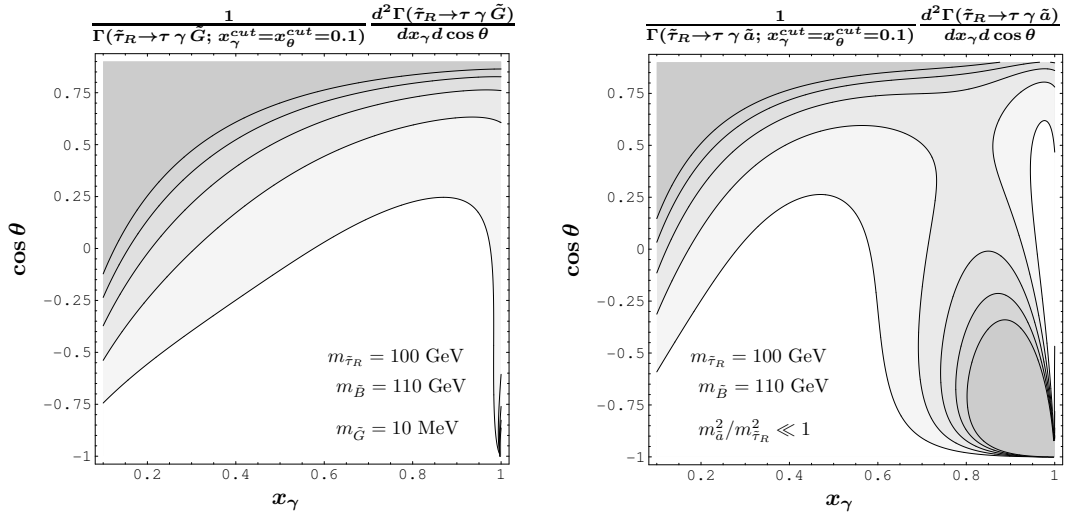}}
\end{picture}
\end{center}
\caption{
The normalized differential distributions of the visible decay products in the 
decays $\widetilde\tau\to\tau\gamma\widetilde G$ for the gravitino LSP 
scenario (left) and $\widetilde\tau\to\tau\gamma\widetilde a$ for the axino LSP 
scenario (right) for $m_{\widetilde\tau_1} = 100\,$GeV, 
$m_{\widetilde B} = 110\,$GeV, $m_{\widetilde a}^2/m_{\widetilde\tau_1}^2 \ll 1$, 
and $m_{\widetilde G} = 10\,$MeV.  
The contour lines represent the values 0.2, 0.4, 0.6, 0.8, and 1.0, where the darker
shading implies a higher number of events. Taken from \cite{Brandenburg:2005he}.}
\label{fig:aG}
\end{figure}
The feasibility of studying 3-body decays and distinguishing gravitinos from axinos
has been discussed in \cite{Brandenburg:2005he}. The distribution of stau decay events
in the two variables $\theta$, the opening angle between the photon and the tau, and 
$x_\gamma\equiv2E_\gamma/m_{\widetilde\tau_1}$ is shown in figure \ref{fig:aG}.
For the gravitino the events are peaked only in the region of soft and collinear
photon emission whereas for the axino a second peak shows up characterized
by a back-to-back tau-photo emission and large photon energies.
For a total number of $10^4$ analyzed stau decays in the scenario considered in
\cite{Brandenburg:2005he} it has been found that $110\pm10$ (stat.) and 
$165\pm13$ (stat.) 3-body decays will be observerd in the gravitno and axino
LSP scenario, respectively, 1\% and 28\% of which are expected to be selected 
by imposing appropriate cuts in the $x_\gamma$-$\cos\theta$-plane.
These numbers illustrate that $\mathcal{O}(10^4)$ of analyzed stau decays
could be sufficient for a significant distinction of those scenarios.

\section{Conclusions}

Supersymmetric scenarios with a very weakly interacting LSP
are well motivated from cosmology. The very weak coupling naturally gives
rise to a long-lived NLSP which is considered to be the lighter stau here. 
These particles usually pass the detector and can be directly detected. 
If these particles will be discovered at the LHC, the ILC/CLIC provides the unique 
environment to study the decays of the 
stau in detail. Reconstructed 2-body decays will
allow for a measurement of the scale of supersymmetry
breaking $\langle F\rangle$ (in the case of a gravitino LSP) or the
Peccei-Quinn scale (in the case of an axino LSP).
From 3-body decays it is even possible to measure the spin of
the LSP. For a gravitino LSP this leads to the attractive
possibility to test the supergravity paradigm. Additionally, the 
measurement of the life-time from 2-body decays provides direct access
to the gravitational coupling. Hence, two independent unequivocal
predictions of supergravity can be probed.


\phantomsection 
\bibliographystyle{../utphys}
\bibliography{../staus}


\end{document}